# Resilience as pathway diversity: Linking systems, individual and temporal perspectives on resilience

Steven J. Lade[1,2], Brian H. Walker[2], L. Jamila Haider[1]

1. Stockholm Resilience Centre, Stockholm University, Stockholm, Sweden
2. Fenner School of Environment and Society, The Australian National University, Canberra, Australia

**ABSTRACT**

Approaches to understanding resilience from psychology and sociology emphasise individuals' agency but obscure systemic factors. Approaches to understanding resilience stemming from ecology emphasise system dynamics such as feedbacks but obscure individuals. Approaches from both psychology and ecology examine the actions or attractors available in the present, but neglect how actions taken now can affect the configuration of the social-ecological system in the future. Here, we propose an extension to resilience theory, which we label 'pathway diversity', that links existing individual, systems and temporal theories of resilience. In our theory of pathway diversity, resilience is greater if more actions are currently available and can be maintained or enhanced into the future. Using a toy model of an agricultural social-ecological system, we show how pathway diversity could deliver a context-sensitive method of assessing resilience and guiding planning. Using a toy state-and-transition model of a poverty trap, we show how pathway diversity is generally consistent with existing definitions of resilience and can illuminate long-standing questions about normative and descriptive resilience. Our results show that pathway diversity advances both theoretical understanding and practical tools for building resilience.

**INTRODUCTION**

In an age of turbulent social and environmental change, the concept of resilience has grown from origins in ecology (Holling 1973) and psychology (Garmezy 1973) to become one of the most popular concepts in sustainable development. Calls are frequently made to build resilience in cities (Meerow et al. 2016), communities (Berkes and Ross 2013) and ecosystems (Chapin III et al. 2009, Mitchell et al. 2014). A large body of research (Folke 2006, Baggio et al. 2015) studies resilience from multiple disciplinary and interdisciplinary perspectives.

Research on social-ecological systems views resilience as "a system's capacity to cope with shocks and undergo change while retaining essentially the same structure and function" (Walker et al. 2009). Resilience may involve adaptation or even transformation to a different kind of system in addition to persistence (Folke et al. 2016). Its historically systems-based perspective has emphasised the roles of feedbacks, nonlinearities and slow and fast variables in generating phenomena such as regime shifts, adaptive cycles, traps and transformations. This systems view of resilience has been critiqued for failing to adequately deal with the roles of individual actors and factors such as power relationships that can limit their agency (Hornborg 2009, Cote and Nightingale 2012, Olsson et al. 2015). While recent research on traps and transformations has begun to deal with these factors (Westley et al. 2013, Boonstra et al. 2016), inclusion of agent-oriented approaches to resilience that incorporate social dimensions of resilience is urgently required (Brown 2016, Cinner and Barnes 2019).

Existing concepts that take a more agent-oriented approach to social-ecological resilience include response diversity and resilience-as-filtering. Originally an ecological concept (Elmqvist et al. 2003), response diversity states that a community (or ecosystem) with a greater variety of responses to a shock will be more likely to withstand that shock (Leslie et al. 2013, Grêt-Regamey et al. 2019). Resilience-as-filtering views resilience as the result of active and passive filtering of old and new practices by actors within a social-ecological system (Haider 2017). However, these existing agent-oriented approaches to resilience are limited in the degree to which they can account for the system feedbacks that have made social-ecological resilience such a powerful concept. For example, a response that degrades a social-ecological system's physical or human assets may not be helpful for the long-term resilience of the system. Furthermore, both systems- and agent-oriented perspectives on resilience focus on resilience at a snapshot in time and have limited capacity to analyse how available responses or system attractors change over time.

Finally, despite the popularity of resilience, it can be unclear how to quantitatively assess resilience or how to use resilience to choose between different policy options. Early metrics of resilience based on distance to a critical threshold (Walker et al. 2004) neither indicate how far from the threshold is desirable nor account for the need to adapt or transform in addition to withstanding change. The seven "principles for building resilience" (Biggs et al. 2015) are useful governance guidelines but do not give guidance on specific policy choices and have limited capacity to analyse the roles of specific actors in building or managing resilience. Instead, economic optimisation is frequently used to guide decision-making, often undermining resilience (Walker and Salt 2006). Formal definitions of resilience that can guide decision-making are needed to better translate resilience into policy and challenge economic optimisation as the dominant decision-making heuristic.

In this Insight article we propose a theory of resilience as pathway diversity (Fig 1) that links systems- and agent-oriented perspectives on resilience. Pathway diversity advances both resilience theory, such as overcoming the persistence paradigm where a poverty trap has high resilience, and application of resilience, such as providing a framework on when to choose between adaptation and transformation. Pathway diversity builds on pathways-based and livelihoods approaches to sustainable development (Scoones 2009, Leach et al. 2010, Wise et al. 2014, Enfors-Kautsky et al. 2018), but focuses on the diversity of available pathways as a measure of resilience. The focus of this manuscript is primarily theoretical, laying the methodological basis for future empirical applications.

## CURRENT APPROACHES TO RESILIENCE

Current understandings of resilience have developed from separate origins in psychology (Garmezy 1973), ecology (Holling 1973) and engineering. Different communities of research and practice maintain at times drastically different conceptions and operationalisations of resilience (Table 1).

### Ecological resilience

In ecology, resilience was originally defined as a "measure of the persistence of systems and of their ability to absorb change and disturbance and still maintain the same relationships between populations or state variables." (Holling 1973). A wide variety of definitions are now used in ecological research and it is not our goal to review them all. Definitions involving the ability to resist and recover from disturbance (Ingrisch and Bahn 2018) are common and are often visualised using the ball and cup diagram (Fig. 2a). Common metrics for measuring resilience include distance to a threshold or more recently the variance of time series (Scheffer et al. 2009).

**Social-ecological resilience**

The recognition that understanding ecological problems requires understanding the behaviour of humans involved with those ecosystems, and vice versa, that humans are dependent on (and indeed shape) ecosystems and other biophysical systems, triggered the development of the field of social-ecological system research (Berkes et al. 2003, Sterk et al. 2017) and the study of social-ecological resilience. Key concepts and heuristics that were inspired by ecosystem dynamics, such as the ball-in-the-cup, the adaptive cycle (Holling 1985), regime shifts, and panarchy (Gunderson and Holling 2002) have been modified and used for social-ecological systems. These concepts and other frameworks emerging from a complex adaptive systems perspective on social-ecological systems are now grouped together under the label of 'resilience thinking' (Walker and Salt 2006, Carpenter and Brock 2008). Resilience thinking has yielded many insights into the dynamics and structure of social-ecological systems (Folke et al. 2016). A key advance was that resilience involves not only persistence, but also involves adaptation in response to disturbances as well as transformation when necessary (Walker et al. 2004, Folke et al. 2010). Social-ecological resilience is also conceptually the broadest of the strands of resilience, seeking to include ecological dynamics, individual human behaviour and collective human behaviour within its analysis (Ostrom 2009).

Apart from the triad of resilience as persistence, adaptability and transformability (Walker et al. 2004), few general conceptual frameworks of what resilience is have been developed to match the concept's transition from ecological to social-ecological systems. A common critique of research on social-ecological resilience from a social science perspective is that it often uses an ontology of systems thinking, inherited from ecological resilience, which has limited its ability to incorporate individual action and constraints on that action such as inequality, power and agency (Cote and Nightingale 2012, Brown 2014, Olsson et al. 2015). First, attractor and ball-and-cup metaphors work best at the system level, aggregating across members of a community or society to create a deterministic picture of the aggregated behaviour of the group. This aggregated, deterministic picture makes it difficult to analyse what actions an individual could have chosen, what factors constrained or enabled those choices, and what inequalities exist within the community or society (Davidson 2010). Second, systems approaches to resilience frequently focus on equilibrium states, which are susceptible to an optimisation or command-or-control treatment, or to an understanding of resilience as merely maintaining the status quo. Third, in system perspectives, a single system structure is usually studied. Changes in resilience over time for example along potentially different development pathways are studied at best implicitly, with the exception of some statistical measures of resilience (Scheffer et al. 2009). While some recent resilience work has addressed these research gaps, for example on transformative agency in social-ecological systems (Westley et al. 2013), social-ecological traps (Boonstra et al. 2016), and sense of place (Masterson et al. 2017), agency is still largely absent from formal frameworks and theories of resilience. An approach that combines systems-oriented and agent-oriented perspectives of resilience would allow agency and power to be dealt with more explicitly (Brown 2016).

From an implementation perspective, current approaches to social-ecological resilience emphasise the complex adaptive systems nature of social-ecological systems and have limited capacity to provide guidance when confronted with a specific policy choice. General guidelines have been offered for how to build resilience (Biggs et al. 2015), global limits have been identified (Steffen et al. 2015) and resilience assessments help understand how a particular social-ecological system operates (Sellberg et al. 2015). There is understandable reluctance to universal measures of resilience, since the features that make social-ecological systems resilient may vary widely (Quinlan et al. 2016) and such a measure could lead to top-down control that actually undermines the resilience of the social-ecological system (Cox 2016). However, the lack of a clear approach to quantifying social-ecological resilience leaves economic optimisation as the dominant quantifiable policy goal. While there exists potential in principle for optimisation methods to be used in conjunction with resilience (Fischer et al. 2009). An approach to quantifying and optimising resilience, which still accounts for context-dependence, could be used to challenge economic optimisation.

**Psychological resilience**

The concept of resilience has as long a history in psychology (Garmezy 1973) as in ecology (Holling 1973). Psychological resilience research studies the ability of individuals to withstand stress, but generally attributes lack of resilience to characteristics of the individual, rather than considering systemic factors. More recently, research into social resilience (Adger 2000) has studied the ability of communities or other groups to withstand disturbance, and systemic and contextual factors are also better taken into account. Psychological, cultural and emotional dimensions of environmental change are increasingly recognised in sustainability science as fundamental in shaping strategies of persistence, adaptation or transformation in the face of uncertainty (Brown et al. 2019). Many of the tools for measuring social resilience have been developed by researchers and practitioners for resilience in development, which we now describe.

**Resilience in development**

Resilience is highly prominent in discussions of international development and poverty alleviation agenda in academia, policy and practice (Brown 2016). Development programming frequently sets 'building resilience' as an explicit goal of interventions, while in the international policy arena resilience is a central concept in frameworks such as the Sustainable Development Goals. Despite the widespread use of the term, there are markedly different understandings about what it is, how it can be assessed (Quinlan et al. 2016), and how it can be built (Béné et al. 2014).

Intensive effort to measure resilience has led to the development of criteria-based resilience metrics (Cinner et al. 2015, FAO 2016). These lists help to incorporate a diversity of views on resilience. However they are so broad that they can be used to justify a wide range of interventions; they neglect the complex interrelationships that comprise poverty; and they fail to be context sensitive, that is, the factors that contribute to resilience in one location may be very different to the factors that contribute to resilience elsewhere (Levine 2014).

In development economic models, the S-curve model of poverty traps is a popular way to understand poverty (Fig. 2c). This S-curve model is conceptually very similar to the multiple stable states in resilience's ball-and-cup heuristic (Barrett and Constas 2014). However the S-curve model leads to a focus on asset inputs that are intended to 'push' individuals or communities over the barrier between poor and non-poor states (Lade et al. 2017). It obscures the capabilities of individuals to act to improve their situation, and the contextual factors such as political or economic situations that may constrain these capabilities (Boonstra et al. 2016). An approach to resilience that deals with these factors would therefore improve theory and practice in poverty alleviation. It would also influence sustainability science more broadly, where traps concepts are increasingly used (Tidball et al. 2016, Haider et al. 2018).

**Pathways conceptions of resilience**

The challenge of navigating along and between different 'pathways to resilience' or 'pathways to sustainability' is a common problem framing. For example, the 'pathways approach' of the STEPS centre (Leach et al. 2010) studies how a particular development pathway can be envisioned and enacted. This perspective emerged from decades of work of the livelihoods approach (Scoones 2009) who proposes to draw on social-ecological resilience theories in order to bring in longer term dynamics into existing livelihoods frameworks which emphasise the diversity of local place-based pathways. The adaptation pathways framework studies sequencing and potential lock-in of adaptation decisions (Haasnoot et al. 2013). Attempts have been made to extend the ball-and-cup model of resilience with a temporal dimension along metaphorical pathways (Enfors 2013, Steffen et al. 2018) (Fig 2b).

Often the implicit view in a pathways perspective is that different pathways lead to different outcomes for resilience and an appropriate pathway should be chosen and moulded carefully. However, this view does not account for the relationship between the number or diversity of different pathways and resilience outcomes. Here, we develop a theory based on a pathways perspective that can incorporate the insights of systems- and individual-oriented approaches of resilience through analysing the diversity of available pathways.

## RESILIENCE AS PATHWAY DIVERSITY

We propose a theory of resilience as pathway diversity, the diversity of future pathways available to an agent or agents (Box 1). In this theory, higher pathway diversity leads to higher resilience. We use a specific understanding of a pathway as a sequence of actions made by an actor or set of actors (Box 1). Decisions that promote resilience, under this theory, are those that maintain existing available actions or improve the array of actions available to actors in the social-ecological system now and into the future.

We now elaborate on the key components of our theory of pathway diversity: constraints limit available actions; greater diversity of actions means greater resilience; but that the consequences of an action on future availability of actions along a pathway must be accounted for. Through these elements, pathway diversity provides a framing that can link individual and systems theories of resilience.

---

Box 1: Definitions

Pathway: A temporal sequence of actions taken by an agent or agents and the associated changes in the social-ecological system in which they are embedded. An agent could be an individual, household, community or other group. In practice, assessment of pathways will need to be truncated at some time horizon.

Pathway diversity: The diversity of pathways available to an agent or agents.

**Constraints on available actions: contributions from individual perspectives on resilience**

Individual or agent-centric perspectives of resilience will be required to understand the factors that enable and constrain different actions, and the pathways to which those actions lead. For example, power relationships could lead to an actor's available actions being suppressed or restricted, or alternatively allow an actor to access previously unavailable actions; the knowledge or physical skills possessed by a person may limit their available choices; habit or preference may constrain available actions; and other available resources such as financial or natural capital may also constrain their actions. There are a number of concepts and frameworks that could be used to study these factors, of which we now give some examples. While consideration of the constraints on agency must be a part of any pathway diversity analysis, we do not here preference any particular analytical framework.

The traps framework of Boonstra et al. (2016) distinguishes between the desires, abilities and opportunities for actors to respond to a trap. Abilities and opportunities provide internal and external, respectively, constraints on an actor's access to actions, while desires enable the actor's capacity to make use of that access. Similarly, the Values-Rules-Knowledge (vrk) framework (Gorddard et al. 2016) distinguishes between the values that determine how likely an actor is to make use of an available action, the rules-in-use and the rules-in-form that constrain or enable the available actions and the knowledge that an actor uses to assess which actions are available.

In development studies, Sen's capabilities approach (Sen 2001) famously laid out the argument that resources (inputs) must be converted into valuable functionings which depends on a person's physical ability, social context and environmental constraints among others. A person's capability is a set of diverse valuable functionings, but the functionings that a person achieves are those that are actually selected by the individual. Our theorisation of pathways diversity therefore only includes pathways on which people actually have the capability to act.

The sustainable livelihoods framework expresses constraints on action through five interlinked capitals (Scoones 1998, 2009, Serrat 2017): financial capital, natural capital, human capital, social capital and physical capital. Levels of these capitals are influenced by environmental, economic and political factors, and levels of these capitals in turn constrain which livelihood options are available. Livelihoods approaches have been noted as being ripe for integration with resilience concepts (Tanner et al. 2015)

**Connecting diversity with resilience**

We connect available actions to resilience by claiming that an actors are more resilient if they have a greater diversity of actions available or, equivalently, a larger 'option space' (Enfors-Kautsky et al. 2018). Diversity is an intuitive surrogate for resilience: the more different things a system has, the better its capacity to respond to disturbances or change (Biggs et al. 2015). For example, biodiversity is believed to be important for the resilience of ecosystems. Response diversity (Elmqvist et al. 2003), which has also been applied to social-ecological systems (Leslie et al. 2013, Grêt-Regamey et al. 2019), identifies diversity of initial response capabilities with resilience. Response diversity is often introduced as a complement to functional redundancy: resilience is improved if multiple entities perform the same function (functional redundancy) and is even further improved if they respond differently to shocks or stresses (response diversity). The diversity of initial responses that is captured by response diversity, however, does not account for the consequences of the initial responses on the longer-term viability of the pathway. Furthermore, while response diversity counts the diversity of species (Elmqvist et al. 2003) or actors (Leslie et al. 2013, Grêt-Regamey et al. 2019) according to their responses, pathway diversity could arise from the different future pathways available to a single agent. Diversity in the pathways of energy flow through an ecosystem has also long been proposed to promote stability (MacArthur 1955); here we also identity diversity of pathways with resilience but do not discuss energy flows.

**Feedbacks from actions: contributions from systems perspectives on resilience**

If an action reduces the actions available to an actor in the future—for example, by running down their natural or financial capital—then that action surely does not contribute to that actor's resilience (Abel et al. 2016). It is therefore important to emphasize that we define resilience not in terms of the diversity of available actions, but rather as the diversity of the available pathways of actions. Available pathways could therefore be enabled or constrained both by feedbacks involving attributes of the agent (such as their abilities or desires) or their environment (such as their biophysical environment or their relationships with other agents) or external drivers (such as the political, economic, social, and biophysical contexts of the actors) (Figure 1). As in structuration theory (Giddens 1986), pathways result from the interplay of systemic feedbacks with individual agency.

In systems language, assessing pathways rather than only initial actions allows the consequences of actions to feed back through their effects on the social-ecological system in which the actor is embedded to affect the actor's future availability of actions. A pathways perspective therefore incorporates the concept of feedback, which is a core element of systems perspectives on resilience. In the section 'Theoretical investigation of pathway diversity: poverty traps' below, we show that pathway diversity is consistent with other systems resilience concepts such as traps and regime shifts. A pathway perspective also moves beyond static understandings of the configuration of a social-ecological system such as the ball-and-cup diagram (Fig 2A) by allowing the possible options of actors in the system to change over time (Fig 2B).

We have not yet stated how the diversity of pathways should be assessed. There are many ways to measure diversity (Stirling 2007). A simple initial approach would be to simply count the number of pathways, out to some specified time horizon. We use this approach in the applied investigation of pathway diversity in the next section. In the section thereafter, we offer a quantitative measure of pathway diversity that allows weights of different pathways to be incorporated. This approach would recognise that pathways that are in principle possible but in practice unlikely to be accessed, for example due to the constraints on actions discussed above, do not significantly contribute to resilience.

**APPLIED INVESTIGATION OF PATHWAY DIVERSITY: AGRICULTURAL RESILIENCE PLANNING**

We now illustrate how pathway diversity can be assessed and used to guide decision-making using a toy model of a farmer in an industrialised, water-stressed society. Toy models, such as the one used here, are deliberately simplistic representations of reality that are useful to succinctly illustrate a concept—here, pathway diversity.

**Identification of pathways**

We assume that the farmer has the following alternative livelihoods:

- Single crop variety monoculture. This farming strategy is highly economically productive but highly susceptible to disease.
- Multiple crop varieties, in which several varieties of the same kind of crop are grown. This strategy means that if disease or financial shocks reduce the viability of one crop, the farmer has other crops from which to choose.
- Mixed cropping, in which several different crops are grown. The farmer can readily convert from growing mixed crops to single crops or single varieties, but not back again.
- Beef farming and cropping. The farmer can readily convert to beef farming or cropping, but not back again without government support for diversification.
- Beef only. We assume that all crops will fail in moderate drought, but beef farming can be maintained. However, in good conditions beef farming is less economically productive than cropping, we assume in this model. In times of severe drought, beef farming will also fail unless supported by drought relief.
- Tourism. If the farmer retains sufficient funds, they can invest in and switch to a livelihood from tourism. Once the farmer has switched to tourism (for example, by selling all farming infrastructure and installing accommodation), the farmer cannot readily switch back to beef or cropping. Tourism is only sustainable if tourism demand is sufficiently strong.
- Exit: The farmer can choose to exit from this set of livelihoods at any time. We do not consider what happens to the farmer after exit. If the farmer persists with non-viable cropping, beef farming or tourism for more than two years, degradation of farmland or loss of financial assets force the farmer to exit.

Under a pattern of various external drivers (Fig 3A), we constructed the pathways available to the farmer. We graphically represented the pathways in a form based on the adaptation pathways framework (Fig 3B). The adaptation pathways framework was developed to aid exploring and sequencing adaptation actions in futures where adaptation options are available for only limited time periods (Haasnoot et al. 2013). It illustrates which alternative actions are available at each point in time from the current state of the social-ecological system.

In the particular pattern of drivers we chose to illustrate this model, disease first causes failure of the single crop strategy. A beef price crash causes exit from the beef-only strategy, but a mixed beef and cropping strategy can be maintained. A moderate drought causes exit from all cropping strategies. This is followed by a severe drought, which would also have caused exit from beef farming were it not for government drought relief. After the end of the drought, diversification assistance can help farmers return to a mixed beef and cropping strategy. Tourism is also vulnerable to shocks, with a reduction in tourism demand leading to exit from tourism.

Wise et al. (2014) identified that adaptation pathways in practice often demonstrate dynamics characteristic of complex systems such as path dependency where a decision affects the availability of future actions, maladaptive spaces that limit future actions, and transformative change where a decision is taken to substantively change the social-ecological system. We can see many of these phenomena in our adaptation pathways (Fig 3B). If cropping is chosen during drought, the farmer becomes locked-in to this livelihood due to feedbacks from declining natural and financial capital and is eventually forced to exit. Lock-in is a form of path dependence (Mahoney 2000, Page 2006) where future decision-making options become substantially narrowed (Allison and Hobbs 2004). We characterise switching between cropping strategies (such as single and multiple varieties) as adaptation. A transition to tourism, however, is a transformation since it involves fundamental re-organisation of the farmer's activity and the farm's infrastructure that cannot readily be reversed.

**Pathway diversity and resilience: guiding decision-making**

We identified the pathways of greatest resilience (green highlights, Fig 3B) in this toy model based on the map of pathways, but not explicitly considering pathway diversity, as follows. Initially, the most resilient livelihood strategy is mixed beef and cropping, since mixed pathways provide the greatest range of future options (Meert et al. 2005). When a moderate drought occurs, however, cropping becomes unviable in this model. Maintaining a cropping strategy would lead to lock-in and then exit within two years (years 6-7). The strategy of highest resilience, which allows the farmer to stay on their land, is to switch to beef farming only once the drought hits and while the option is still available (Year 6). Drought relief helps the farmer withstand severe drought (Year 9), while diversification support allows the farmer to return to a high-resilience mixed cropping and beef strategy. This pathway allows the farmer to deal with known threats while also leaving as many options as possible open for unknown threats, that is, to maintain both specific and general resilience (Folke et al. 2010).

We now demonstrate that pathway diversity predicts pathways of greatest resilience that are consistent with the intuitive understanding above. For all livelihood strategies at all points in time, we assumed that the farmer had full knowledge of future pathways up to 2 years in the future. We calculated pathway diversity by counting the number of distinct action pathways starting from that livelihood strategy out to 2 years in the future. For example, if the farmer is engaged in mixed cropping coming in to Year 2, they have 10 decision pathways available for the coming two years (Years 2-3): Mixed cropping-Mixed Cropping; Mixed cropping-Multiple varieties; Multiple varieties-Multiple varieties; Mixed cropping-Tourism; Mixed cropping-Exit; Multiple varieties-Tourism; Multiple varieties-Exit; Tourism-Tourism; Tourism-Exit; and Exit-Exit. Each year's actions with maximum pathway diversity (Fig 3C, bolded entries) correspond exactly with the intuitively chosen pathways of greatest resilience (Fig 3B).

We have shown that pathway diversity offers an operational definition of resilience that is readily computed from an adaptation pathways diagram. Whereas conventional approaches to resilience recognise the need for persistence, adaptation and transformation (Folke et al. 2016), a pathway diversity approach recommended a transformation at exactly the same time as intuitive understandings of resilience. Pathway diversity as implemented here could therefore guide decision-making and policy-making to build resilience.

The toy model we have used here has many limitations. (a) Each of the livelihoods considered here and the ease of switching between them are likely subject to many more factors. Transformation to tourism as livelihood will require specific knowledge and skills and may require substantial financial capital. (b) Full knowledge of pathways up to two years in the future is fully known by farmers. In practice, decisions are constrained by uncertainty about the future trajectories of specific drivers and deep uncertainty about unknown future pressures. An operational version of this framework will have to be probabilistic, with probabilities assigned to possible future drivers. Further research is needed that will apply pathway diversity to more realistic situations or models that better reflect the dynamics of real social-ecological systems and the uncertainties associated with decision-making. This research could include embedding a pathway diversity analysis within a participatory resilience assessment and planning process.

# THEORETICAL INVESTIGATION OF PATHWAY DIVERSITY: POVERTY TRAPS

The agricultural example in the previous section investigated how a pathway diversity framework could be used to empirically assess resilience and guide decision making. We now explore assessment of resilience within a mathematical model. A systematic mathematical operationalisation of pathway diversity would allow it to be formally compared with existing systems approaches with resilience, and also open the possibility for a quantitative assessment of resilience as pathway diversity in mathematical models. In this section, we offer causal entropy as a metric of pathway diversity and apply this to a toy model of a poverty trap.

**Causal entropy: Mathematical implementation of pathway diversity**
A likely candidate for implementation is the mathematical formalism of causal entropy (Wissner-Gross and Freer 2013). Causal entropy is a measure of the diversity of future pathways that are accessible from a specified starting state within a specified time horizon. The causal entropy of state $x_i$ in a discrete state space with time horizon $\tau$ is

$$S(x_i, \tau) = -\sum_j \sum_k P(x_{jk}, t+\tau | x_i, t) \log P(x_{jk}, t+\tau | x_i, t) \qquad (1)$$

where $j$ and $k$ denote the different pathways available to endogenous and exogenous degrees of freedom and $P(x_{jk}, t+\tau | x_i, t)$ are the probabilities of the system taking the different pathways denoted by $j$ and $k$. It is causal because all pathways causally connected to an initial state $x_i$ are considered, and an entropy because it measures the unpredictability of which pathway will be taken. We caution that the use of the term entropy here is purely in a descriptive sense, as a measure of the diversity of pathways, as it is used in information theory. Physical laws associated with thermodynamic entropy, such as the second law of thermodynamics stating that entropy must increase in a closed system, have no relevance here. Confusion between the different varieties of entropy used in different branches of physics has led to substantial misuse and misunderstanding (Kovalev 2016).

Causal entropy was first widely used in astrophysics (Brustein and Veneziano 2000). Wissner-Gross and Freer (2013) applied causal entropy to mathematical models intended to reproduce intelligence tests involving tool use. They controversially claimed that intelligent agents, or complex systems in general, tend to follow a path generated by causal entropic forcing. Again, we avoid any such prescriptive ambitions, but rather use causal entropy as a descriptive measure of pathway diversity.

The mathematical definition of causal entropy (Eq. 1) allows for two sources of variability: endogenous and exogenous to the system under study. In the exogenous case, the freedom of any actors within the system is ignored, and variability emerges from the possible actions of agents or entities external to the system. 'Agent-less' models of resilience, such as distance to a threshold, are compatible with the exogenous view: while there are decisions being made that affect the distance to a threshold, these decisions are not explicitly incorporated in the model. The endogenous case, however, provides an opportunity for agency to be incorporated into a model setting. In the endogenous case, agency can be attributed to actors internal to the system. Constructing a model that displays causal entropy therefore requires the possible actions of agents within the system—such as navigating away from a threshold—to be made explicit. Limitations to an individual's agency could be specified according to factors including power relationships, norms, available assets and individual motivation.

**Descriptive and prescriptive resilience: Pathway diversity analysis of a poverty trap**

We illustrate the potential of causal entropy as a measure of resilience using a simple state-and-transition model (Westoby et al. 1989, Bestelmeyer et al. 2017) of a poverty trap (Fig 4A). The model consists of three states: an initial state A; a state B from which recovery to A usually occurs; and a third state C, reachable from B with low probability, from which recovery to A and B is possible but unlikely. C corresponds to a poverty trap, A corresponds to a 'non-poor' state with greater freedom, and the transition B to C corresponds to a regime shift into the poverty trap.

We calculated the causal entropy of each of the states A, B and C out to a time horizon of 10 transitions by brute force: mapping all possible pathways of length 10 and probabilities associated with all those paths. The results show that the trapped state C has the lowest causal entropy, since there are few pathways available from that state, while state A has highest causal entropy (Fig 4B). The causal entropy of A and B are even higher when there are multiple pathways for returning to A from B and for remaining at A (Fig 4B, dotted lines).

Resilience as pathway diversity might therefore be able to shed light on long-standing discussions about descriptive and prescriptive versions of resilience (Béné et al. 2014, Olsson et al. 2015). Resilience researchers usually maintain that resilience is a descriptive concept. For example, under the conventional association of resilience with persistence, both the trapped state C and the non-poor state A would have high resilience as they are easily able to maintain their state (and C would have the highest resilience), but the resilience of A would be labelled a 'good' resilience while the resilience of C would be labelled a 'bad' resilience (Béné et al. 2014). Outside academia, however, resilience is often used normatively or prescriptively: resilience is always good, therefore A would be high resilience but C low resilience. In normative language high 'bad' resilience is sometimes referred to as rigidity.

Interpreting resilience as pathway diversity, however, gives A as the highest resilience state and C as the lowest resilience state. Pathway diversity is therefore a descriptive measure of resilience that, unlike stability metrics, assesses a poverty trap as a state of low resilience. Pathway diversity could therefore be seen as an implementation of more recent social-ecological approaches to resilience, for example, that define resilience as the capacity to adapt or transform (Folke et al. 2016), which would also assess the trapped state C as low resilience.

We therefore find that pathway diversity is consistent with existing understandings of resilience as persistence that associate proximity to a regime shift as a reduction in resilience, but also matches with more modern notions of resilience as capacity to adapt or transform that associate a poverty trap with low resilience. We also show that resilience as capacity to recover from shocks is enhanced if there are more options available to deal with different kinds of shocks.

Maximising pathway diversity in a model setting could offer an alternative to economic optimisation as a policy goal. While fully maximising pathway diversity may also be undesirable, as we argue below, it could nevertheless be used to illuminate alternative policy goals. The final policy may be a choice between economic optimisation and maximum pathway diversity depending on stakeholder priorities.

## DISCUSSION

We take this opportunity to anticipate potential critiques of pathway diversity as resilience. First, in introducing our theory of resilience as pathway diversity, we do not aim to replace any existing theories or tools for resilience. Rather, we see pathway diversity as a complementary concept that can unite previously disparate approaches to resilience and help guide decision-making for resilience. For example, we have shown that pathway diversity can guide decision-making in a way that: distinguishes when it is better to adapt or transform; links individual constraints on decision-making with systemic feedbacks; is consistent with classical resilience concepts associating regime shifts with a loss of resilience; and is consistent with more modern understandings of a poverty trap as a state of high persistence (rigidity) but low resilience. General guidelines for resilience such as the seven principles for building resilience (Biggs et al. 2015) contribute factors that improve the number and accessibility of pathways. Resilience assessments contribute to revealing key components and critical thresholds at which certain options lead to significant impacts on the social-ecological system (Walker et al. 2009, Sellberg et al. 2015, 2018, Enfors-Kautsky et al. 2018) and key assets or capacities that contribute to a system's resilience (FSIN 2014, FAO 2016).

Second, pathway diversity is a quantification of resilience and therefore could be used to design policies to maintain or increase resilience; indeed, we used it to calculate the pathways of greatest resilience in the toy agricultural examples above. Resilience researchers are justifiably wary of general metrics of resilience, for example because it could lead to top-down control to the detriment of the system that is being managed (Cox 2016), or out of suspicion that any general metric exists that could be usefully applied across the wide variety of contexts and systems in which resilience is used (Quinlan et al. 2016). To these concerns we note that to assess pathway diversity requires a full description (or as full as is available) of the social-ecological system, including alternative options and the consequences of options expressed via feedbacks. Any assessment of pathway diversity will therefore be highly specific to the system being assessed. While we must remain attentive to risks associated with quantifying resilience, pathway diversity presents an opportunity to capitalise on the possible synergies between optimisation approaches and resilience (Fischer et al. 2009) within the factors that constrain resilience in a specific context.

Third, we claim that pathway diversity as resilience is descriptive, not normative. In the poverty trap model, we showed that the pathway diversity of a poverty trap is low, in line with understandings of a poverty trap as a situation of low resilience. Poverty traps are also undesirable and therefore low resilience aligns with low desirability in this case. High pathway diversity, however, is not necessarily good. For example, high pathway diversity and thereby high resilience of a tyrant would be undesirable for much of the country's population (though the pathway diversity of the members of that population might be low). At a more detailed level, one might also ask: Is having access to many 'bad' pathways better than a few 'good' pathways? Our definition of pathway diversity shows that if the pathways are 'bad' because they reduce availability of future actions in some way, then many bad pathways will likely have less pathway diversity than a few good pathways. If, however, good and bad pathways have access to similar number of future actions, then many bad pathways will have higher pathway diversity than a few good pathways.

Lastly, while resilience as pathway diversity generally increases when there are more actions available, it does not promote maximising available actions at any cost. If some actions are harmful to the broader social-ecological system, in that they decrease the availability of future actions, then these actions contribute little to pathway diversity and resilience. Likewise, if the cost of maintaining availability of options is accounted for, then maintaining a large array of options may decrease resilience due to reducing the financial capital needed to access those actions. Resilience could then be seen as the task of balancing navigation along a desirable path with maintaining a large "search space" of alternative options (Prokopenko and Gershenson 2014).

There remains substantial work to develop tools for application of pathway diversity both in quantitative modelling and in resilience planning. In the analyses above, we used a computationally intensive approach to compute causal entropy as pathway diversity. For more models with more alternative states, in particular models with a continuous state space, more efficient methods will be needed to compute and to choose pathways that maximise pathway diversity. Our scenarios were also highly stylised. Further development of this pathway diversity approach to analysing resilience calls for two kinds of activities. First, whether pathway diversity is a useful framing for resilience should be tested in resilience planning workshops. Second, methods to deal with uncertainties in future trajectories of external drivers, future social-ecological dynamics, and current and future options of actors will be needed if pathway diversity is to be a practical tool for decision-making.

## CONCLUSION

We have proposed a theory of resilience called pathway diversity that views resilience as the diversity of current actions and capability to maintain or enhance the diversity of options in the future. Pathway diversity links existing individual and systemic theories on resilience by viewing available actions as subject to social-ecological feedbacks from past decisions. Pathway diversity moves beyond static representations of social-ecological systems such as the ball-and-cup to represent how the available options and trajectories in a social-ecological system change over time. We illustrated how pathway diversity could be used in resilience planning, where we showed that pathway diversity matched an intuitive prediction of the pathway of highest resilience amongst adaptation and transformation options. We also illustrated how pathway diversity can be used to analyse quantitative models, where we showed that pathway diversity is consistent with and builds upon existing approaches to resilience. While there remains substantial work to refine the pathway diversity concept and further develop tools for its application, we believe that it contributes towards resolving the diverse definitions of resilience and to translating the resilience concept into practice.

**Table 1: Summary of resilience definitions and metrics from various fields.** The definitions given are representative only; within each field there may be strong debate and a large diversity of views on resilience.

| Field | Representative definition | Metrics | Weaknesses |
|---|---|---|---|
| Ecology | "the persistence of systems and …their ability to absorb change and disturbance" (Holling 1973) | Distance to threshold Size of basin of attraction Eigenvalue Variance | Originally did not recognise that resilience sometimes requires adaptation or transformation |
| Social-ecological systems | "a measure of a system's capacity to cope with shocks and undergo change while retaining essentially the same structure and function." (Walker et al. 2009)(Folke et al. 2016) | 'Resilience principles' (Biggs et al. 2015) offer guidelines for policy. | General concept that can be difficult to implement. |
| Psychology | "The ability to bounce back from negative emotional experiences." (Tugade et al. 2004) | Sets of individual criteria (Rodriguez-Llanes et al. 2013) | Attributes resilience to the individual; does not recognise systemic factors. |
| Development | "Capacity of a person, household or other aggregate unit to avoid poverty in the face of various stressors and in the wake of myriad shocks." (Barrett and Constas 2014) | Sets of indicators of individual capabilities | Metrics are not context-sensitive (Levine 2014). A single resilience measure may block deeper understanding of system dynamics (Quinlan et al. 2016) |

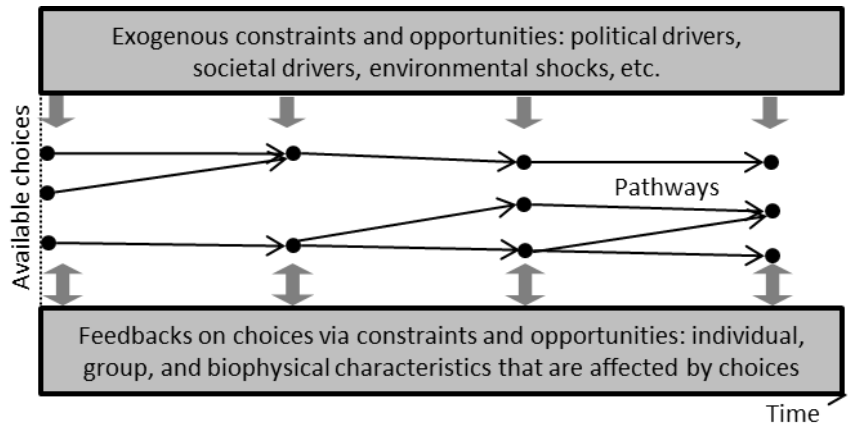

**Figure 1: Pathways and pathway diversity.** The diversity of future pathways are constrained and enabled by exogenous and endogenous drivers and endogenous feedbacks

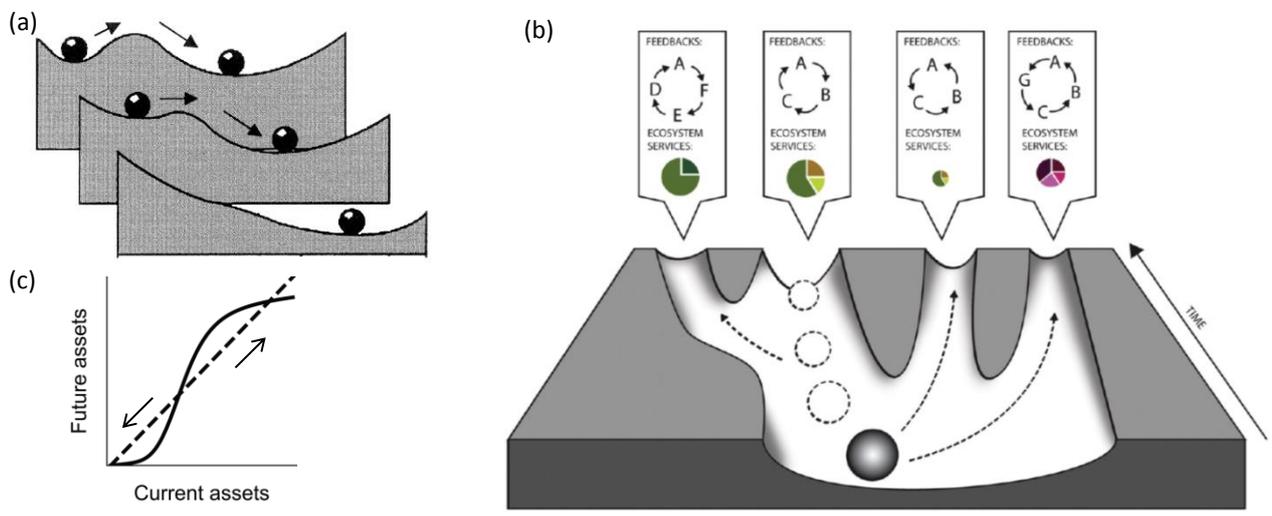

**Figure 2. State-of-the-art resilience theoretical frameworks.** (a) Ball-and-cup diagrams displaying a regime shift. Stable states occur at the bottom of the landscape's 'valleys'. Reproduced from Gunderson (2000). (b) Expanding the ball and cup diagram to trajectories of ecosystem services. Reproduced from Enfors (2013). (c) The S-curve of economic poverty trap models. A fixed point occurs when future assets is the same as current assets, that is, when the solid line crosses the dashed diagonal line; this S-curve shows a stable 'poor' and 'non-poor' states separated by an unstable state. Modified from Lade et al. (2017).

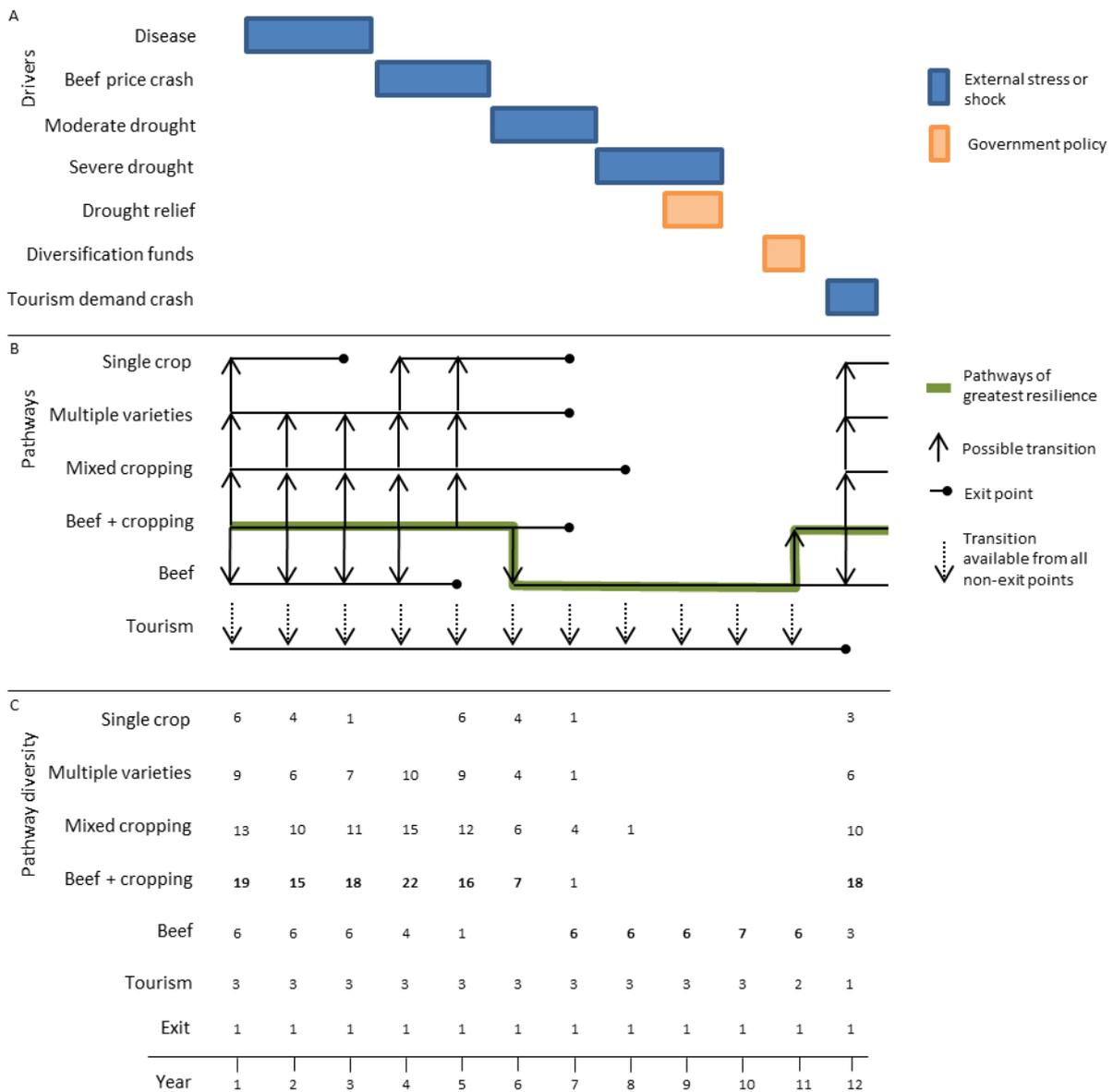

**Figure 3.** Pathway diversity analysis of agricultural decision making. (A) Drivers that influence the available options. (B) Available pathways. (C) Calculation of pathway diversity for each actions at every point in time. The actions of maximum pathway diversity in each year are bolded.

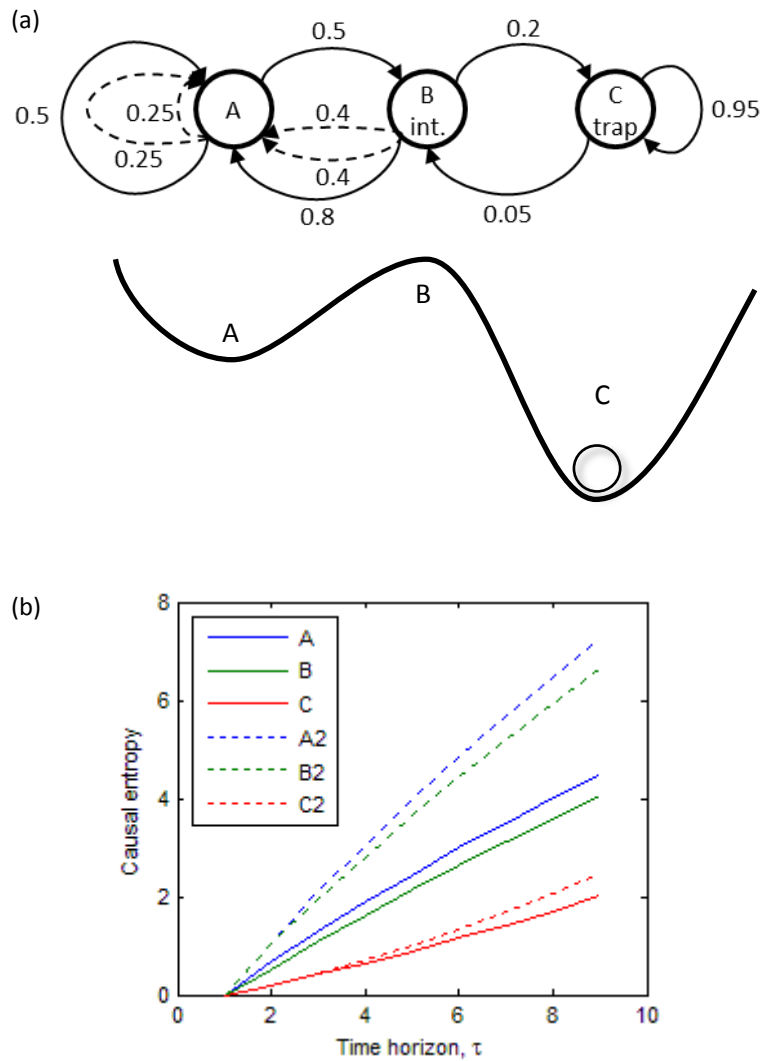

**Figure 4: Pathway diversity of a poverty trap.** (a) State-and-transition model showing a state A which recovers from a transition to state B with high probability (0.8), but if a transition to C occurs then recovery is highly unlikely (0.05). The intention, as illustrated in the ball-and-cup diagram, is that C corresponds to a trapped state of high persistence; A is a more desirable stable state with less persistence, and B is an unstable intermediate state. (b) Causal entropy for the different states as a function of the time horizon. A is the state of highest causal entropy, followed by B, then the trapped state C. Where there are two recovery paths from B to A as well as two paths to maintain A, the causal entropy of states A and B are even higher (dotted lines, A2-C2 in figure legend)